\documentclass[aps,preprint]{revtex4}
\usepackage{amsmath}
\usepackage{epsfig}
\usepackage{float}
\usepackage{dcolumn} 
\usepackage{braket}

\begin{document}
\title{\bf Real discrete spectrum of complex PT-symmetric scattering potentials}  
\author{Zafar Ahmed$^1$, Joseph Amal Nathan$^2$, Dhruv Sharma$^3$, Dona Ghosh$^4$}
\affiliation{$~^1$Nuclear Physics Division, Bhabha Atomic Research Centre, Mumbai 400 085, India \\
$~^2$Reactor Physics Design Division, Bhabha Atomic Research Centre, Mumbai 400085, India \\
$~^3$ National Institute of Technology, Rourkela, 769008 India \\
$~^4$Astavinayak Society, Vashi, Navi-Mumbai, 400703, India} 
\email{1:zahmed@barc.gov.in, 2:josephan@barc.gov.in, 3-sharmadhruv@gmail.com, 4: rimidonaghosh@ gmail.com}
\date{\today}
\begin{abstract}
\noindent
We investigate the parametric evolution of the real discrete spectrum of several complex PT symmetric scattering potentials of the type $V(x)=-V_1 F_e(x) + i V_2 F_o(x), V_1>0, F_e(x)>0$ by varying $V_2$ slowly. Here $e,o$ stand for even and odd parity and $F_{e,o}(\pm \infty)=0$. Unlike the case of Scarf II potential, we find a general absence  of the recently explored accidental (real to real) crossings of eigenvalues in these scattering potentials. On the other hand, we find  a general presence of coalescing of real pairs of eigenvalues to the complex conjugate pairs at a finite number of exceptional points. We attribute such coalescings of eigenvalues  to the presence of a finite barrier (on the either side of $x=0$ ) which has been linked to a recent study of stokes phenomenon in the complex PT-symmetric potentials. 
\end{abstract}
\maketitle

The discovery [1] that complex PT-symmetric Hamiltonians may have real discrete spectrum has given rise to PT-symmetric quantum mechanics.
The coalescing of real discrete eigenvalues to complex conjugate eigenvalues has been known earlier as a phenomenon of spontaneous breaking of complex PT-symmetry. The exactly solvable complex PT-symmetric Scarf II scattering potential
\begin{equation}
V_S(x)=-V_1 ~ \mbox{sech}^2x+i V_2 ~ \mbox{sech}x ~\tanh x, \quad V_1, V_2 \in {\cal R}, V_1>0,
\end{equation}
has been well known to display spontaneous breaking of PT-symmetry  when [2] $V_2=V_{2c}=1/4+V_1 \ (2m=1=\hbar^2)$. In the theory of  exceptional points (EPs) of non-Hermitian potentials 3], the value(s) of $V_2=V_{2c}$ are called EPs where the the real pairs of eigenvalues coalesce and just after the turn into  complex conjugate pairs.  Exactly at these values the corresponding eigenstates become linearly dependent and the Hamiltonian looses diagonalizability.

A recent study of the Stoke's phenomenon of complex PT-symmetric potentials claims [4] the occurrence of level-coalescing (they call it level-crossings) at infinite exceptional points in the potential $V(x)=ig(x^3-x)$. In Ref. [4], this potential is called "PT-symmetric 
double well" with "two wells at $x=\pm 1/\sqrt{3}$". Such potentials as having this feature have been claimed to have infinite number level-coalesings. However, in more simple terms this $V(x)$ is such that its imaginary part has a finite barrier on the either side of $x=0$ according as $g$ is positive or negative. In this work, we show that evolution of several complex PT-symmetric scattering potentials 
whose imaginary part has a barrier on the either side of $x=0$ have a finite number of level-coalescings at the critical values $V_2=V_{2c}$.  It is important to recall that the complex PT-symmetric potentials $V(x)=x^2+igx, V(x)=igx^3, V(x)=-V_1 \mbox{sech}^2x+iV_2 \tanh x$ [5] do not entail coalescing of levels and exceptional points. However, the interesting potential $V(x)= x^4+igx$ [6] does have them.

Recently, it has been found that  the potential (1) has a very interesting property wherein real discrete eigenvalues cross at special values of $V_{2*}$. This phenomenon has been called accidental crossing [7] of real discrete eigenvalues in one-dimension. As in one dimension, the degeneracy (two (distinct )linearly independent eigenstates having coincident eigenvalue) cannot occur consequently the crossing levels have linearly dependent eigenstates. This  gives rise to loss of diagonalizability of the Hamiltonian which in turn hampers the completeness of the spectrum of the potential.

Interestingly, the solvable regularized one-dimensional complex harmonic (RCHO) oscillator [8,9] potential also had this feature of level-crossings however this had come up more clearly in two recent Refs. [10,11].

In the parlance of exceptional points of non-Hermitian Hamiltonians the above-mentioned two types (real to real, real to complex)  of crossings  of levels may not  be distinguished. However the real to real crossing of eigenvalue appears to be so rare, that so far only two  potentials RCHO [8,9] and Scarf II (1) have yielded it. Curiously, the former is only binding (infinite spectrum) potential that does not allow scattering whereas the latter allows both  bound (finite spectrum) and scattering states. In this work, we also wish to bring out the non-occurrence of accidental level crossings in complex PT-symmetric scattering potentials yet they look (Fig. 1) like Scarf II.

So, with the motivation of studying level coalescings and level-crossings we propose to find the parametric evolution of the finite number of eigenvalues for five models of complex PT-symmetric scattering potentials (See Fig. 1) employing various methods.

\begin{figure}[ht]
\centering
\includegraphics[width=7. cm,height=5. cm]{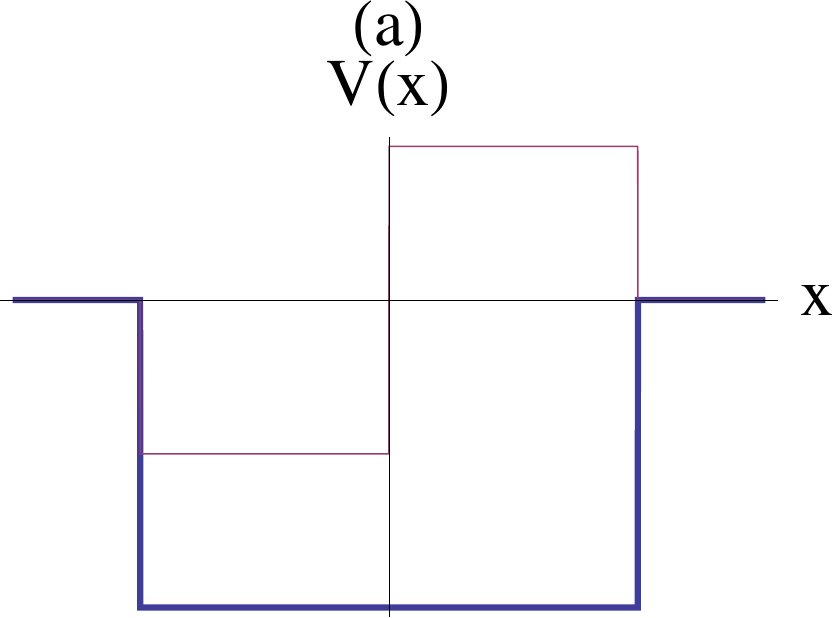}
\hspace{.5 cm}
\includegraphics[width=7. cm,height=5. cm]{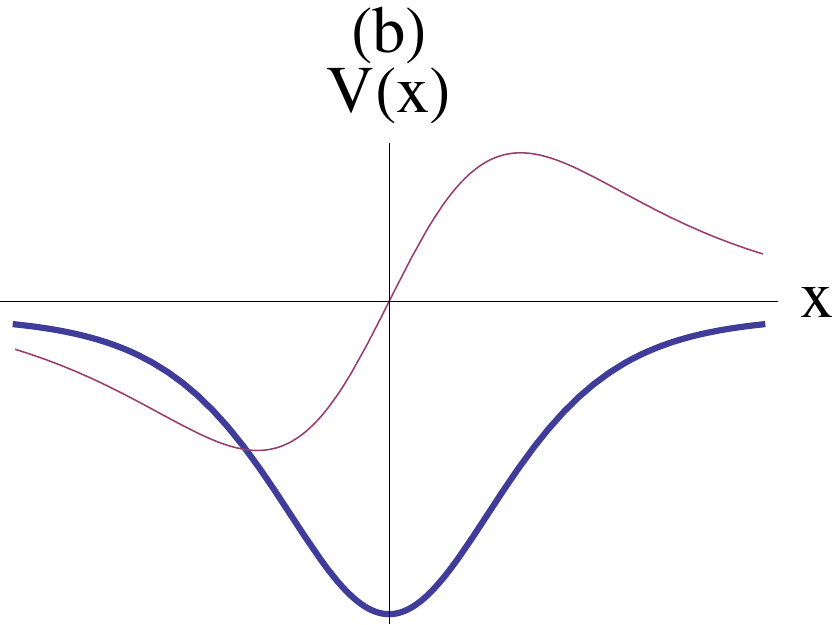}
\caption{Schematic depiction of the Complex PT-symmetric scattering potentials, real part (thick line), imaginary part (thin line). (a) represents the rectangular well $V_R(x)$ (4) and (b) represents $V_S(x) (1), V_G(x)(5),V_F(x) (6), V_H(x)(7)$ and $V_{WC}(x)$ (14)}
\end{figure}

\begin{figure}[h]
\centering
\includegraphics[width=18 cm,height=16. cm]{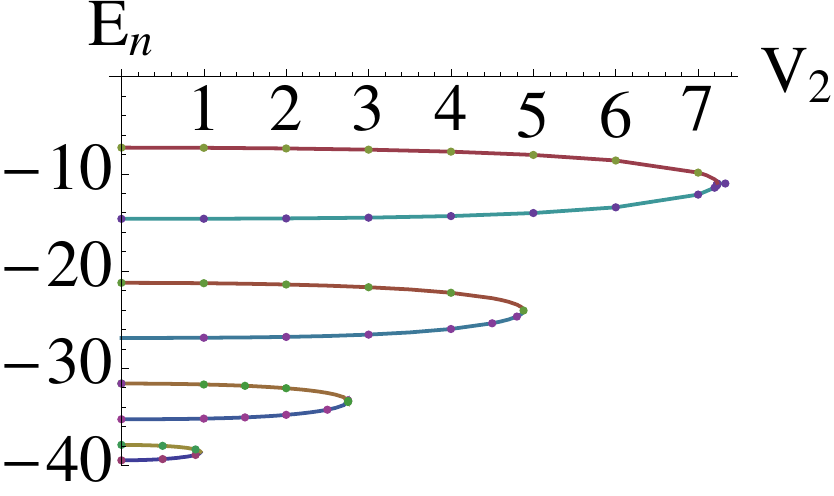}
\caption{The parametric evolution of real discrete spectrum for reatangular potential with $V_1=40$ and $a=2$, notice that there are no level-crossings, but the eigenvalue pairs do coalesce at the exceptional points $V_2=0.96,2.75,4.88,7.33.$. The solid line is due to numerical integration using (3) and the dots are due to the exact analytic method using Eq. (9)}
\end{figure} 
We shall be solving the one-dimensional time-independent Schr{\"o}dinger equation
\begin{equation}
\frac{d^2\psi(x)}{dx^2}+\frac{2m}{\hbar^2} [E- V(x)]\psi(x)=0
\end{equation}
for five models of one-dimensional complex PT-symmetric scattering potentials. These potentials (See Fig. 1) essentially vanish at $x=\pm\infty$ and their real part constitutes a well which support only a finite numbers of real discrete eigenvalues. Their imaginary parts are the corresponding anti-symmetric profile.These potential well  are piece-wise constant rectangular, Gaussian, $-(1+x^4)^{-1}$, $-\mbox{sech} x$ and Wigner-Coulomb profiles.

We would like to outline our method of finding real discrete eigenvalues by numerical integration of the Schr{\"o}dinger equation (2). Let us define $k={\sqrt{2\mu(-E}}/hbar$. We take the general solution of (2) as $\psi(x<-L)= C e^{kx},\psi(-L<x<L)= A u(x) + B v(x),\psi(x>L)= D e^{-kx}$. Here $u(x)$ and $v(x)$ are linearly independent solutions of (2), their initial values as $u(0)=1,u'(0)=0$  and $v(0)=0, v'(0)=1$ to start the integration to both left and right side up to $-L$ and $+L$, respectively. $L$ is sufficiently large distance to be chosen. We match  these piece-wise solutions and their first derivatives at $x=-L,0,L$. Finally, we eliminate $A,B,C,D$ in the resulting equations to obtain the eigenvalue formula
\begin{equation}
\frac{ku(L)+u'(L)}{kv(L)+v'(L)}=\frac{ku(-L)-u'(-L)}{k v(-L)-v(-L)}
\end{equation}
to find the eigenvalues. 

We choose the distance $L$ such that the final results (eigenvalues) have the desired accuracy. For all the calculations here we use $2\mu=1=\hbar^2.$ Using (3), we fix a value of $V_1$ so that there are at least 6 real discrete eigenvalues for the real potential well $(V_2=0)$ by varying $E=-V_1$ to $E=0$.
Next, $V_2$ is proposed to vary slowly till we get real pairs of eigenvalue which curves coalesce to complex conjugate pairs of eigenvalue. We call these special values of $V_{2}$ as $V_{2c}$ which are known as exceptional points (EPs) of non-Hermitian potential. Nonetheless, we are interested to see whether or not there will be crossings of real discrete eigenvalues from real to real when we vary  $V_2$. The eigenvalues of complex PT-symmetric potentials (4,5,6,7,14) considered here in the sequel satisfy $E_n(-V_2)=E_n(V_2)$, we therefore evaluate $E_n(V_2)$ only for $V_2>0$.

First, to confirm our numerical method,  we take up the well known rectangular complex PT-symmetric profile [12] of width $2a$.
\begin{equation} 
V_R(x)=-V_1 \Theta_1(x)-iV_2 \Theta_2(x), 
\Theta_1(x)=\left\lbrace\begin{array}{lcr}
1, & &  |x|\le a\\
0, & &  |x|>a\\
\end{array}
\right.,
\Theta_2(x)=\left\lbrace\begin{array}{lcr}
0, & & |x|\ge a\\
-1, & & -a < x <0\\
1, & & 0 \le x <a \\
\end{array}
\right.
\end{equation}
which is also  solvable analytically (see Eq. (9) below).
The potential (4) being of finite support we take $L=a=2$. In fig.1, the solid lines  are due to numerical integration method using Eq. (3). No crossing (real to real) of eigenvalues is observed but eigenvalues coalesce at $V_2=V_{2c}=0.96,2.75,4.88,7.33$.  Next, we find the evolution of Gaussian model for $V_1=50$
\begin{equation}
V_G(x)=V_1 e^{-x^2}+ i V_2 xe^{-x^2}, \quad V_1, V_2 \in {\cal R}, V_1>0.
\end{equation}
By this method a distance of $L=10-12$
has been found sufficient for the convergence of eigenvalues. In Fig. 2, the solid lines represent the result due to Eq. (3). The exceptional points for this potential (5) are $V_2=V_{2c}= 43.26, 55.55,63.70$ and the crossing of levels is not found. In Fig. 4,  the parametric evolution of the spectrum of 
\begin{equation}
V_F(x)=\frac{-V_1}{1+x^4}+\frac{iV_2x}{1+x^4},\quad V_1, V_2 \in {\cal R}, V_1>0
\end{equation} 
is presented. The pairs of eigenvalues are well separated from each other and the levels do not cross. They do coalesce to complex conjugate pairs at $V_2=V_{2c}=19.39,38,87,46.35$, we have fixed $V_1=50.$ So, for $V_2<19.39$ all the real discrete eigenvalues are real and PT-symmetry is un-broken. After this critical value, the initial discrete eigenvalues start disappearing. Let us now consider s sech-hyperbolic potential
\begin{equation}
V_{H}=-V_1 \mbox{sech}x +i V_2 \mbox{sech}x \tanh x,\quad V_1, V_2 \in {\cal R}, V_1>0
\end{equation}
whose real part is $\mbox{sech}x$ unlike the Scarf II potential (1). See in Fig. 5, the eigenvalue pairs are well separated without crossing each other, they do coalesce at $V_2=V_{2c}=  25.37, 31.15, 34.92, 37.35$.

In the following, we  find eigenvalues of (4)  alternatively  by the exact and analytic method which will also confirm the results in Fig. 2. We insert this potential in the Schr{\"o}dinger equation (2). Assuming $2m=1=\hbar$, we define
\begin{equation}
p=a \sqrt{E+V_1-iV_2}, \quad q=a\sqrt{E+V_1+iV_2},\quad r=ak, \quad k=\sqrt{-E}.
\end{equation}
The solution of (1) for this potential can be written as $\psi(x<-a)=F e^{kx}, \psi(-a<x<0)=C\sin qx+ D \cos qx, \psi(0<x<a)= A \sin px +B \cos qx, \psi(x>a) =G e^{-kx}.$ By matching these solutions and their first derivative at $x=-a,0,a$, we eliminate $A,B,C,D,F,G$ to get the eliminant as
\begin{equation}
2pqr\cos p \cos q + p(r^2 - q^2)\cos p \sin q + q(r^2 - p^2)\sin p\cos q -r(p^2 + q^2)\sin p \sin q =0,
\end{equation}
which serves as an analytic eigenvalue equation to be solved by varying $E$ from $-V_1$ to 0. By fixing $V_1=20$, and $a=2$, we obtain the parametric evolution of the spectrum of (4). These results are shown by dots in Fig. 2. See an excellent agreement between the two.
\begin{figure}[h]
\centering
\includegraphics[width=18 cm,height=16. cm]{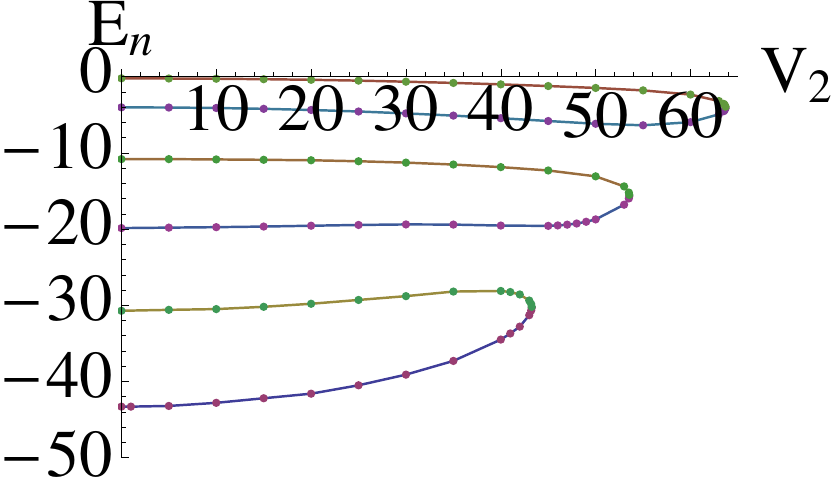}
\caption{The same as in Fig. 2, for $V_G(x)$ (5). The solid line is due to Eq. (3) and dots are due to Eq. (13). The exceptional values are $V_2=43.26, 55.55,63.70.$} 
\end{figure} 
\noindent

Below, we propose to find the eigenvalues of the Gaussian potential (5) potential alternatively by  the diagonalization of $H=p^2/(2\mu)+V_G(x)$ in the harmonic oscillator (HO) basis.
For HO basis $\ket{n}$, we know that $H_0=-\frac{d^2}{dx^2}+x^2, H\ket{n}=(2n+1)\ket{n}$, where $2\mu=1=\hbar=\omega$. Using the well known $a,a^{\dagger}$ operators, we know that
\begin{equation}
\braket{m|p^2|n}=-\frac{\sqrt{(n-1)n}}{2}~\delta_{m,n-2}+\frac{(2n+1)}{2}~\delta_{m,n} - \frac{\sqrt{(n+1)(n+2)}}{2}~\delta_{m,n+2}.
\end{equation}
More interestingly the following required matrix elements can be found analytically with help of available but rare integrals [13]
\begin{equation}
\braket{m|e^{-x^2}|n}= \cos[(m-n)\pi/2] \frac{\Gamma[(m+n+1)/2]}{\sqrt{2 \pi m!n!}}, 
\end{equation}
and 
\begin{equation}
\braket{m|xe^{-x^2}|n}=\left\lbrace\begin{array}{lcr}\frac{(m-n)}{2\sqrt{2}\sin[(m-n)\pi/2]}\frac{\Gamma[(m+n)/2]}{\sqrt{2 \pi m!n!}}, \quad \mbox{if}, & & m+n=\mbox{odd} \\
0, & & \mbox{otherwise}.
\end{array}
\right.
\end{equation}

\begin{figure}[h]
\centering
\includegraphics[width=18 cm,height=16. cm]{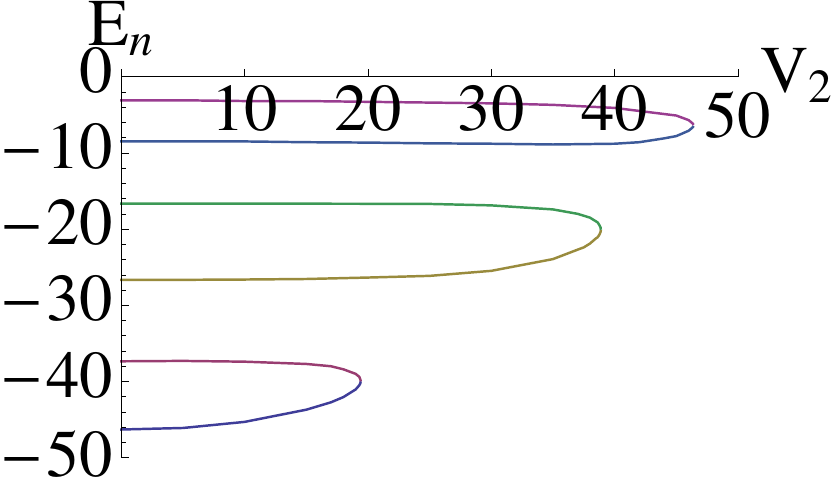}
\caption{The same as in Fig. 2, for $V_F(x)$ (6). These results are due to Eq. (3) and the exceptional point are $V_2= 19.39,38,87,46.35$
and $V_1=50$.}
\end{figure}

Using (10,11,12), we write the matrix elements $h_{m,n}=\braket{m|H|n}$ to get the eigenvalues as
\begin{equation}
\det|h_{m,n}-E ~\delta_{m,n}|=0.
\end{equation}   
In fig. 3, See the excellent agreement between solid lines (using Eq. (3) and dots (obtained by diagonalization using Eq. (13)).

Our last model to be discussed is the Wigner-Coulomb type complex PT-symmetric scattering potential expressed as
\begin{equation}
V_{WC}(x)=\frac{-V_1}{1+x^2}+ \frac{i V_2 x}{1+x^2}, \quad V_1,V_2 \in {\cal R}, V_1>0.
\end{equation}
The Schr{\"o}dinger equation for this potential is  known to be un-amenable. A special case $(V_1=V_2)$ of this potential is $V_{WC}(x)=\frac{iV_2}{x-i}$ which is a complex regularized PT-symmetric potential and on the real line $x\in (-\infty,\infty)$ its discrete spectrum is null. This may well be understood by realizing that for any real value of $E$ it gives rise to only one classical turning point (for bound states there should at least be two turning points). 
This special case also serves to {\it a priori}  indicate that $V_{2c}<V_1.$  The  aforementioned special case of complex Coulomb potential has been treated [14,15] on a special complex trajectory to find a real discrete spectrum. 

In order to solve the eigenvalue equation $H\psi=E\psi$, where $H=[\frac{p^2}{2\mu}+V(x)]$. Here we use a special method [16,17] wherein we find the roots of $E$ by solving the determinantal equation
\begin{equation}
\det|<m|(1+x^2)H-E x^2)|n>-E\delta_{m,n}|=0,
\end{equation}
where $|n>$ are the well known harmonic oscillator eigenstates.
The motive behind choosing this method [16,17] is two-fold. The matrix elements like $<m|(1+x^2)^{-1}|n>$ cannot be found analytically.
As the  determinant becomes larger and larger, the analytic
matrix elements are more desirable. Secondly, the imaginary part of this potential (1) like the Coulomb potential varies as $\sim 1/|x|$, asymptotically. This insufficiently rapid asymptotic fall off of the coulomb potential brings in the typical  problems in the  integration of the Schr{\"o}dinger for asymptotic values.  We take $2\mu=1=\hbar^2$, the eigenvalue equation (2)  can be expressed as 
\begin{figure}[h]
\centering
\includegraphics[width=18 cm,height=16. cm]{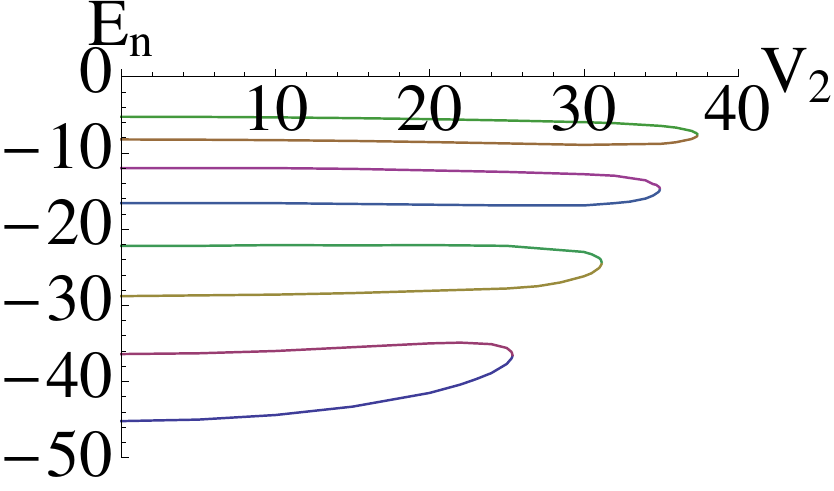}
\caption{The same as in Fig. 2, for $V_H(x)$ (7), $V_1=50$. The exceptional points are $V_2=25.37, 31.15, 34.92, 37.35$} 
\end{figure} 

\begin{eqnarray}
&\det|{\cal H}_{m,n}(E)|=0 \nonumber\\
&{\cal H}_{m,n}=\braket{m|p^2|n}+\braket{m|x^2 p^2|n}-(E+V_1)\braket{m|n}+iV_2 \braket{m|x|n}-E\braket{m|x^2|n}.
\end{eqnarray}

\begin{equation}
\braket{m|x|n}=\sqrt{\frac{n}{2}}~\delta_{m,n-1}+ \sqrt{\frac{(n+1)}{2}}~\delta_{m,n+1}
\end{equation}
\begin{equation}
\braket{m|x^2|n}=\frac{\sqrt{(n-1)n}}{2}~\delta_{m,n-2}+\frac{(2n+1)}{2}~\delta_{m,n}+ \frac{\sqrt{(n+1)(n+2)}}{2}~\delta_{m,n+2}
\end{equation}
\begin{eqnarray}
&\braket{m|x^2 p^2|n} = -\sqrt{(n-3)(n-2)(n-1)n} ~ \delta_{m,n-4}/4 + \sqrt{n(n-1)}  \delta_{m,n-2} + 
\nonumber \\
&(2n^2+2n-1) ~ \delta_{m,n} 
-\sqrt{(n+1)(n+2)} ~ \delta_{m,n+2}/4
-\sqrt{(n+1)(n+2)(n+3)(n+4)} 
\delta_{m,n+2}/4\nonumber \\
 &- \sqrt{(n+1)(n+2)(n+3)(n+4)}  \delta_{m,n+4}/4.
\end{eqnarray}
Curiously, the parametric evolution of eigenvalues obtained for this potential is qualitatively different from the other ones: $V_R(x), V_G(x), V_F(x), V_H(x)$ discussed above. See in Fig. 6, the initial eigenvalue curves are longer which go on becoming shorter for higher eigenvalues. Once again  there are no crossings of eigenvalues, but eigenvalues coalesce at $V_2=19.73, 10.87, 5.53, 2.75$.
\begin{figure}[h]
\centering
\includegraphics[width=18 cm,height=16. cm]{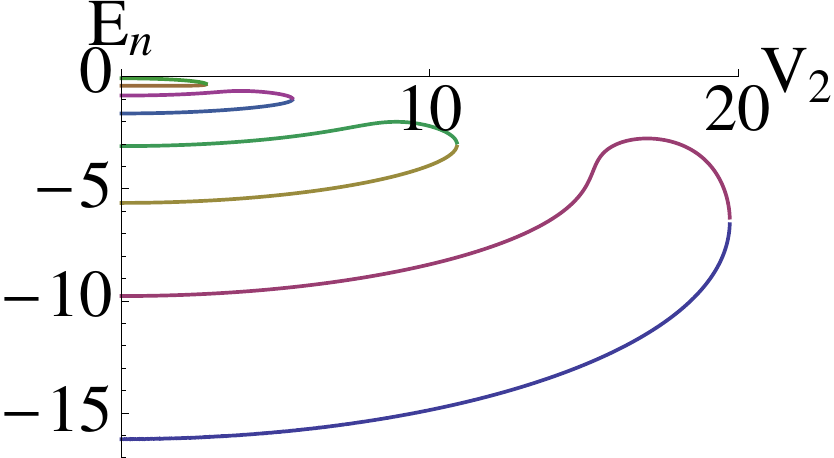}
\caption {The same as Fig. 2, for the Wigner-Coulomb potential (14), $V_1=20, a=2$. The exceptional points are $V_2=19.73,10.87,5.53,2.74$. These results are due to the special method [12,13] of diagonalization (16)} 
\end{figure} 

To conclude, in this paper we have found parametric evolution of eigenvalues for five complex PT-symmetric scattering potentials employing different methods. Such works are instructive and desirable especially when complex PT-symmetry makes way in to  textbooks of quantum mechanics. Unlike the Scarf II, these scattering potentials do not give rise to the real to real crossing of discrete eigenvalues. In this regard, the present work also attracts the attention  on Scarf II (and also the RCHO) as two  highly exceptional cases of accidental crossings of eigenvalues. The reason behind their specialty needs to be investigated.  However, the PT-symmetric scattering potentials discussed here do display a general coalescing of eigenvalues at a finite number exceptional points, we attribute  
this phenomenon to the occurrence of a finite barrier either side of $x=0$. This may be sufficient but not a necessary condition on a potential for the occurrence of level-coalescing. In this, regard further confirmations and investigations are welcome. 
 

\section*{\Large{References}}
\begin{enumerate}
\bibitem{1} C.M. Bender and S. Boettcher,  Phys. Rev. Lett. {\bf 80} (1998) 5243. C.M. Bender, Rep. Prog. Phys. {\bf 70} (2007) 947.
\bibitem{2} Z. Ahmed, Phys. Lett. A  {\bf 282} (2001) 343; {\bf 287} (2001) 295.
\bibitem{3} T. Kato, `Perturbation Theory of Linear operators`, Springer, N.Y. Springer (1980).`
\bibitem{4} R. Giachetti and V. Grecchi, `Level crossings in a PT-symmetric double well', arxiv: 1506:0167[math-ph].
\bibitem{5} G. Levai and E. Magyari, J,. Phys. A: Mat. Theor. {\bf 42} (2009) 195302.
\bibitem{6}C.M. Bender, M.V. Berry, O.N. Meisinger, V. M. Savage and M. Simsek, J. Phys. A : Math. Gen. {\bf 34} (2001) L31. 
\bibitem{7} Z. Ahmed. J. A. Nathan, Dona Ghosh, Gaurang Parkar, Phys. Lett. A {\bf 327} (2015) 2424.
\bibitem{8} M. Znojil, Phys. Lett. A {\bf 259} (1999) 220.
\bibitem{9} B. Bagchi, C. Quesne, and M. Znojil, Mod. Phys, Lett. A {\bf 16} (2001) 2047.
\bibitem{10} M. Znojil arXiv 1303:4876, D.I. Borisov, Acta Polytech. {\bf 54} (2014) 93, arXiv 1401:6324,
\bibitem{11} "Non-Selfadjoint Operators in Quantum Physics: Mathematical Aspects", John Wiley $\&$ Sons, Inc.; July 2015,ISBN 978-1-118-85528-7
(edited by Fabio Bagarello, Jean-Pierre Gazeau, Franciszek H. Szafraniec and Miloslav Znojil) Fig. on p.22. 
\bibitem{12} M. Znojil, Phys. Lett. A {\bf 285} (2001) 7.;
Z. Ahmed, Phys. Lett. A {\bf324 }(2004) 154;
Z. Ahmed, J. Phys. A: Math. Teor.  {\bf 47} (2014) 385303.
\bibitem{13} A. Erdelyi, W. Magnus, F. Oberhettinger, F.G. Tricomi,
{\em Table of Intregral Transforms} Vo. II (McGraw-Hill: New York, 1954) pp.289-290.
\bibitem{14} M. Znojil and G. Levai, Phys. lett. A {\bf 271} (2000) 327.
\bibitem{15} G. Levai, Pramana j. Phys. 6{\bf 73} (2009) 329.
\bibitem{16} K.V. Bhagwat, J. Phys. A Math. Gen. {\bf 14} (1981) 377.
\bibitem{17} S. J. Hamerling, {\em Latent Roots and Latent Vectors} (London: Adam Hilgr,1954) pp. 62-70.
\end{enumerate}
\end{document}